\documentclass{article}

\usepackage{arxiv}

\usepackage[utf8]{inputenc} % allow utf-8 input
\usepackage[T1]{fontenc}    % use 8-bit T1 fonts
\usepackage{hyperref}       % hyperlinks
\usepackage{url}            % simple URL typesetting
\usepackage{booktabs}       % professional-quality tables
\usepackage{amsfonts}       % blackboard math symbols
\usepackage{nicefrac}       % compact symbols for 1/2, etc.
\usepackage{microtype}      % microtypography
\usepackage{lipsum}
\usepackage{graphicx}% Include figure files
\usepackage{dcolumn}% Align table columns on decimal point
\usepackage{bm}% bold math
\usepackage{bbm}% bold math

\usepackage{amsmath}
\usepackage{amssymb}
\usepackage{verbatim}

\title{Bootstraps Regularize Singular Correlation Matrices}
\author{Christian Bongiorno\\
Universit\'e Paris-Saclay, CentraleSup\'elec,\\ Laboratoire de Math\'ematiques et Informatique pour les Syst\`emes Complexes,\\ 91190, Gif-sur-Yvette, France}

\begin{document}

\maketitle

\begin{abstract}
I show analytically that the average of $k$ bootstrapped correlation matrices rapidly becomes positive-definite as $k$ increases, which provides a simple approach to regularize singular Pearson correlation matrices. 
If $n$ is the number of objects and $t$ the number of features, the averaged correlation matrix is almost surely positive-definite if $k> \frac{e}{e-1}\frac{n}{t}\simeq 1.58\frac{n}{t}$ in the limit of large $t$ and $n$. The probability of obtaining a positive-definite correlation matrix with $k$ bootstraps is also derived for finite $n$ and $t$. Finally, I demonstrate that the number of required bootstraps is always smaller than $n$. This method is particularly relevant in fields where $n$ is orders of magnitude larger than the size of data points $t$, e.g., in finance, genetics, social science, or image processing.
\end{abstract}

\section{Introduction}

Correlation and covariance matrices are fundamental dependence estimators in statistical inference. Their use includes risk minimization in finance~\cite{markowitz1959portfolio}, analysis of  functional genomics~\cite{schafer2005shrinkage}, or image processing~\cite{velasco2015comparative}.  However, when the number of objects under study ($n$) exceeds the number of available data points ($t$), theses matrices cannot be inverted. As a result, many standard inference methods cannot be applied directly. To overcome this issue, a large literature on eigenvalue regularization has been devoted to this issue over the last decades. 
%Some noise filtering approaches use Random Matrix Theory to correct the eigenvalue distribution so as to reduce the noise effect of the finite sample size. The most popular ones are eigenvalue clipping~\cite{laloux1999noise} and IW-regularization~\cite{bun2017cleaning}. Unfortunately, these two methods can be applied only in the $t>n$ regime.
 The most relevant ones are the Ledoit-Wolf linear shrinkage~\cite{ledoit2004well} and the more recent non-linear shrinkage~\cite{ledoit2017nonlinear}. These methods, apart from regularizing singular correlation matrices, attempt to reduce the noise effect due to finite sample size.  In addition, Ref.~\cite{higham2002computing} proposes a recursive algorithm that aims to find the most similar positive-definite matrix to an initial problematic matrix that is not positive-definite. Similarly to the proposed method, this approach does not try to denoise the target matrix but corrects the eigenvalue distribution by removing the non-positive eigenvalues.

In this work, I propose a simple alternative approach based on bootstrap resampling to regularize correlation matrices with $z>0$ zero degenerate eigenvalues. In particular, I prove that the probability to obtain a positive defined matrix from the average of $k$ bootstrap resampling scenarios converges rapidly with respect to $k$ to one provided that $k$ is larger than $\frac{e}{e-1} \frac{n}{t}$.

\section{The Bootstrap Average Correlation Matrix}
Let $\textbf{X} \in \mathbb{R}^{n \times t}$ be the data matrix and $\textbf{C} \in \mathbb{R}^{n \times n}$ its Pearson correlation matrix. We assume that no column or row of $\textbf{X}$ is a linear combination of the others; this implies that $\textbf{C}$ has rank $r = \min \{ n, t-1\}$.  Let $\textbf{X}^{(b)} \in \mathbb{R}^{n \times t}$ be a bootstrap copy  of $\textbf{X}$ obtained by sample replacement of the columns of $\textbf{X}$, and $\textbf{C}^{(b)}$ its correlation matrix. A generic element of $\textbf{X}^{(b)}$ is $x^{(b)}_{ij} = x_{i h^{(b)}_j}$, where ${\bf{h}}^{(b)}$ is a vector of dimension $t$ obtained by random sampling with replacement of the elements of vector $(1, 2, \cdots, t)$.

This paper derives an approximate expression of the probability that the smallest eigenvalue $\lambda_0$ of the correlation matrix $\langle \textbf{C} \rangle := k^{-1} \sum_{i=0}^k \textbf{C}^{(i)}$ is larger than  zero as a function of the number of bootstrap copies. The minimum number of bootstrap copies $k^+$, that guarantees $\langle \textbf{C} \rangle$ to be positive-definite within a chosen confidence level, shows a real transition in the large-system limit, defined here as $n,t\to\infty$ at fixed $q$. 
%% P(lmd) hass a transistion. Define  thermodynamic limit transition

\section{The Distribution of the Number of Null Eigenvalues}
The first step is to obtain a probability distribution of the number of zero eigenvalues $z_b$ of a given bootstrap correlation matrix $\textbf{C}^{(b)}$. One has
\begin{equation}\label{eq:z}
z_b = \max \{ n + 1 - u_b, \, 0 \},
\end{equation}
where $u_b$ is the number of unique column indices sampled from $\textbf{X}$ in the $b$-bootstrap copy. The exact probability distribution of $u_b$ is known to be~\cite{mendelson2016distribution} 
\begin{equation}\label{eq:s2k}
\mathcal{P}(u_b) = \frac{\mathcal{S}_2(t,u_b)\, t!}{t^t \, (t-u_b)! }
\end{equation}
where $\mathcal{S}_2(t,u_b)$ is the Sterling number of the second kind. Such a distribution has mean and variance
\begin{equation}\label{eq:exactmom}
\left.\begin{aligned}
\mu(t) &= t \, \left[ 1 - \left( 1 - \frac{1}{t} \right)^t \right]  \\
\sigma^2(t) &= t \, \left( 1 -  \frac{1}{t} \right)^t + t^2\, \left( 1-  \frac{1}{t} \right) \left( 1-  \frac{2}{t} \right)^t  - t^2\, \left( 1-  \frac{1}{t} \right)^{2t} .
\end{aligned}\right.
\end{equation}
In the limit of large $t$,
eqs~\eqref{eq:exactmom} become
\begin{equation}\label{eq:approxmom}
\left.\begin{aligned}
\mu(t)  &\approx  \left(1-\frac{1}{e}\right)t + \frac{1}{2 e} \\
\sigma^2(t) & \approx   \left(\frac{e - 2}{e^2}\right) t  +\frac{3 -e}{2 e^2}\,.
\end{aligned}\right.
\end{equation}
\begin{figure}[t]
\centering
\includegraphics[width=0.45\columnwidth]{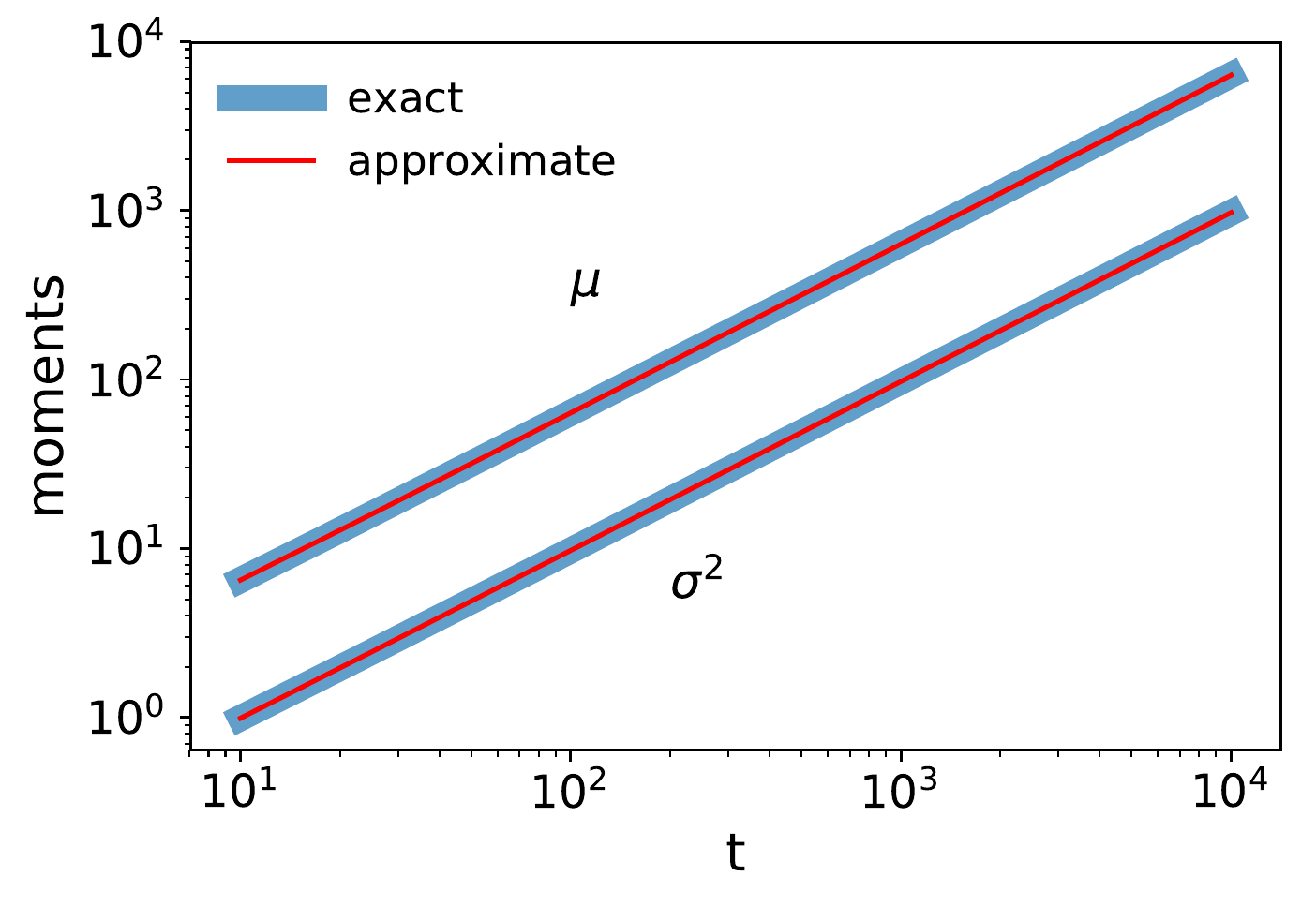} 
\includegraphics[width=0.45\columnwidth]{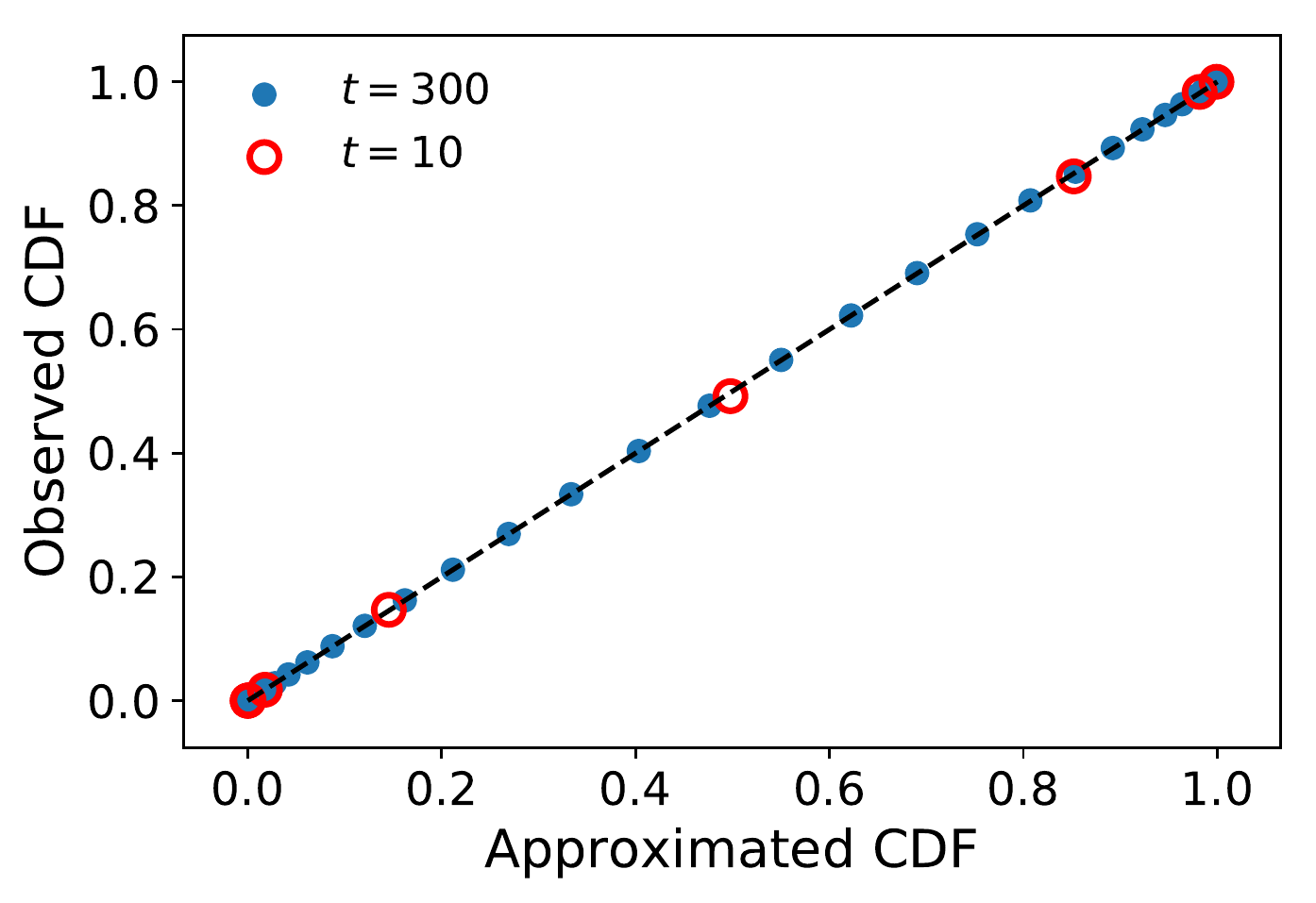} 
\caption{The left plot shows the exact (Eq.~\eqref{eq:exactmom}) and approximate (eq.~\eqref{eq:approxmom}) $t$-dependence of the first two moments of $\mathcal{P}(u_b)$ distribution of Eq.~\eqref{eq:s2k}. The right plot shows the approximate Normal Cumulative Distribution Function (CDF) against the observed CDF obtained with $10^4$ random sampling for every integer value of $u_b \in [0,t]$. }\label{fig:moments}
\end{figure}

Furthermore, it is worth noticing that the deviation of the empirical $\mathcal{P}(u_b)$ from a normal $\mathcal{N}(\mu(t),\sigma(t))$ is negligible for even for moderately large $t$~\cite{mendelson2016distribution} , as reported in the right-hand side plot of Fig.~\ref{fig:moments}.

If we consider a condition characterized by an abundance of expected zero eigenvalues, i.e., $n \gg t$, then the probability distribution of $z_b$ according to Eq.~\eqref{eq:z} can be approximate by a Normal distribution
\begin{equation}\label{eq:norm}
\mathcal{P}(z_b) \approx \mathcal{N}\left(n+1 -\mu(t) , \sigma(t) \right) \,.
\end{equation} 

Now that the distribution of the zero eigenvalues for the single bootstrap copy is known, we can answer the original question, and consider $k$ bootstrap copies of $\textbf{X}$ such that $\langle \textbf{C} \rangle := k^{-1}\sum_{i=1}^k \textbf{C}^{(i)}$.

To make further progress, it is necessary to recall the geometrical properties of the space associated to degenerate eigenvalues.  Let us suppose that $\textbf{C}^{(i)}$ has $z_i$ zero eigenvalues. Then the set of eigenvectors associated with these zero eigenvalues defines a hyper-plane $V_i$ of dimension $z_i$ embedded in an $n$ dimensional space. Each vector $\textbf{w}$ that lies in $V_i$ verifies $\textbf{w}\, \textbf{C}^{(i)} \, \textbf{w}' = 0$; however, if there is at least another $j\ne i$ whose $z_j$ zero eigenvalues of $\textbf{C}^{(j)}$ define hyper-plane $V_j$ such that $\mbox{dim}(V_i \cap V_j) \ngeq 1$, then $\textbf{w}\, \textbf{C}^{(j)} \, \textbf{w}' > 0$;  and thus $\textbf{w}\, \langle \textbf{C} \rangle \, \textbf{w}' > 0$ for every vector $\textbf{w}$ that lies in $V_i$ or $V_j$.

It is important to point out that eigenvectors associated to $z_i$ zero eigenvalues can be assumed to be ``randomly'' chosen with the constraint to be orthogonal with $V_i^{\perp}$, the space defined by the eigenvectors associated with the $n-z_i$ non-zero eigenvalues; this because they do not carry any information about the correlation matrix $\textbf{C}^{(i)}$ since they explain zero variance. Therefore every rotation of the basis of $V_i$ constrained to be orthogonal with $V_i^\perp$ will produce exactly the same matrix $\textbf{C}^{(i)}$. In the $k=2$ case, the probability that $\mbox{dim}(V_1 \cap V_2) \ngeq 1$ will be approximately $1$ if $z_1 + z_2 \leq n$ and $0$ otherwise. It is possible to visualize this relationship easily in a three-dimensional space, i.e., $n=3$. In case of two random straight lines, that have dimensions $z_1=1$ and $z_2=1$, the probability that they intersect in a straight line is almost zero since they must be coincident;  differently, if we consider two random planes $z_1=2$ and $z_2=2$ they will intersect in a straight line almost surely apart from only configurations in which they are parallels. The above-discussed approximation, in the case of the spectral decomposition, is valid if the probability that the orthogonal spaces $V_1^\perp$ and $V_2^\perp$ defined from the $n - z_1$ and $n - z_2$ non-zero eigenvalues perfectly overlap is negligible. In a bootstrap resampling, when $t$ is sufficiently large, this probability is approximately zero, as this requires to sample the same column indices of $\textbf{X}$ for both bootstrap realizations, in other words, $\textbf{C}^{(1)}=\textbf{C}^{(2)}$. 

More generally, for $k$ bootstrap copies, every hyper-plane $V_i$  will verify $\mbox{dim}(V_i \cap V_j) \ngeq 1$ for at least one $j \neq i$ with probability $1$ if
\begin{equation}\label{eq:cond}
\zeta := \sum_{i=1}^k z_i \leq (k-1) \,n .
\end{equation}
If the above inequality holds, then $\langle \textbf{C} \rangle$ has no zero eigenvalue.
From Eq.~\eqref{eq:cond}, one can derive an upper bound for the number of bootstrap copies required. In fact, even if all bootstrap correlations have $n-1$ null eigenvalues, no more than $k = n$ bootstrap copies are necessary to obtain a positive define matrix $\langle \textbf{C} \rangle$. 
 
According to Eq.~\eqref{eq:norm}, the distribution of $\zeta$ can be approximated by a sum of $k$ identical normal distributions that converges to
\begin{equation}\label{eq:Pxi}
\mathcal{P}(\zeta) \approx \mathcal{N}\left( k\,(n+1-\mu(t)), \sqrt{k}\,\sigma(t)\right).
\end{equation}
Therefore, the probability that the smallest eigenvalue $\lambda_0$ of $\langle \textbf{C} \rangle$ is larger than zero can be obtained from the cumulative distribution function of $\mathcal{P}(\zeta)$ estimated at $(k-1)\,n$, that is
\begin{equation}\label{eq:Pl}
\mathcal{P}(\lambda_0 > 0 ) \approx \mathcal{P}\left(\zeta \leq (k-1)\,n \right) =  \int^{(k-1)\,n}_{-\infty} \mathcal{P}(\zeta)\, d\zeta  \approx  \frac{1}{2} \left[ 1 + \mbox{erf}\left(\frac{ [\mu(t)-1] \, k - n}{ \sigma(t)\, \sqrt{2 k}  } \right) \right]
\end{equation}

\begin{figure}[htb]
\centering
\includegraphics[width=0.45\columnwidth]{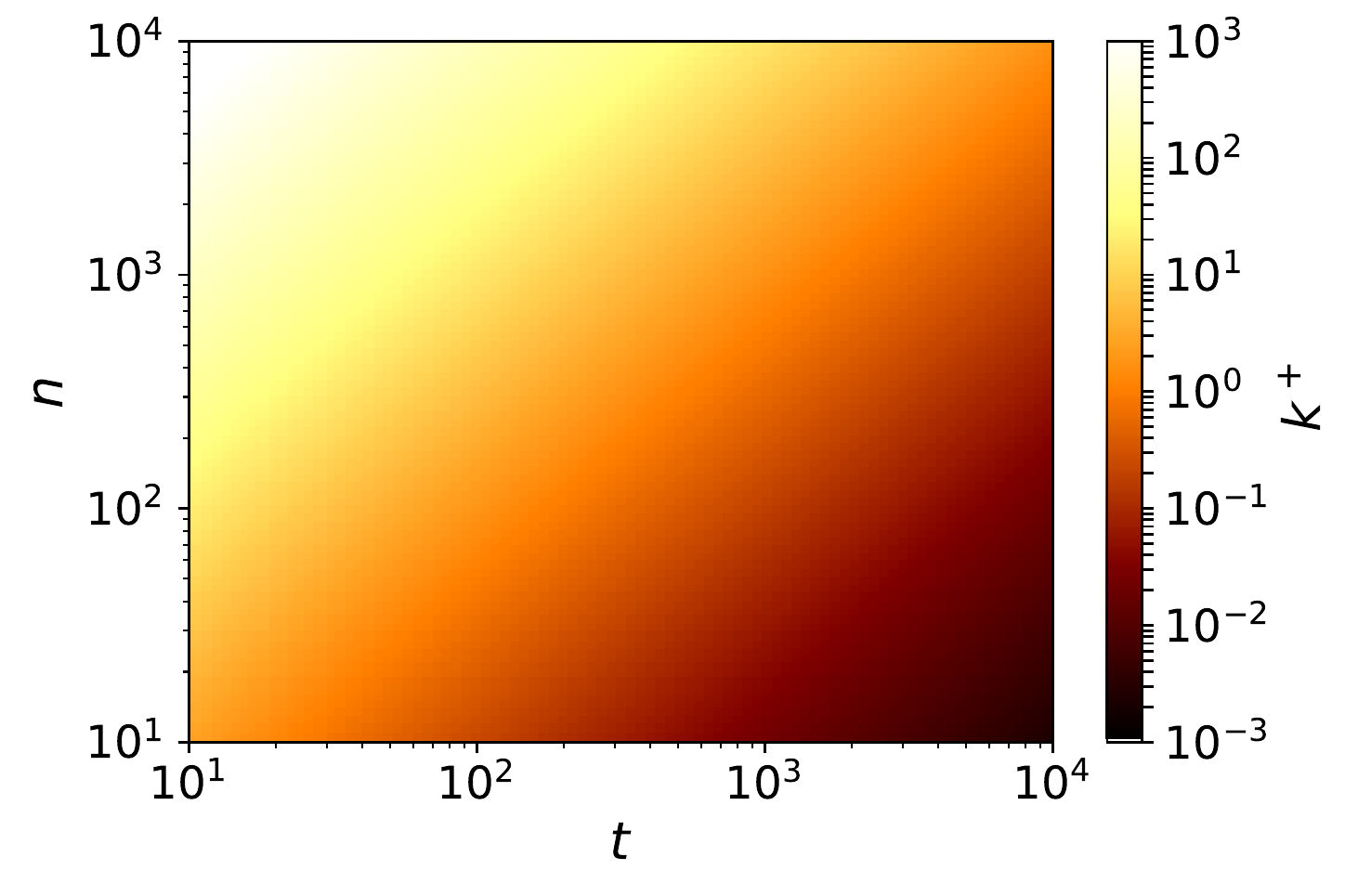} 
\includegraphics[width=0.45\columnwidth]{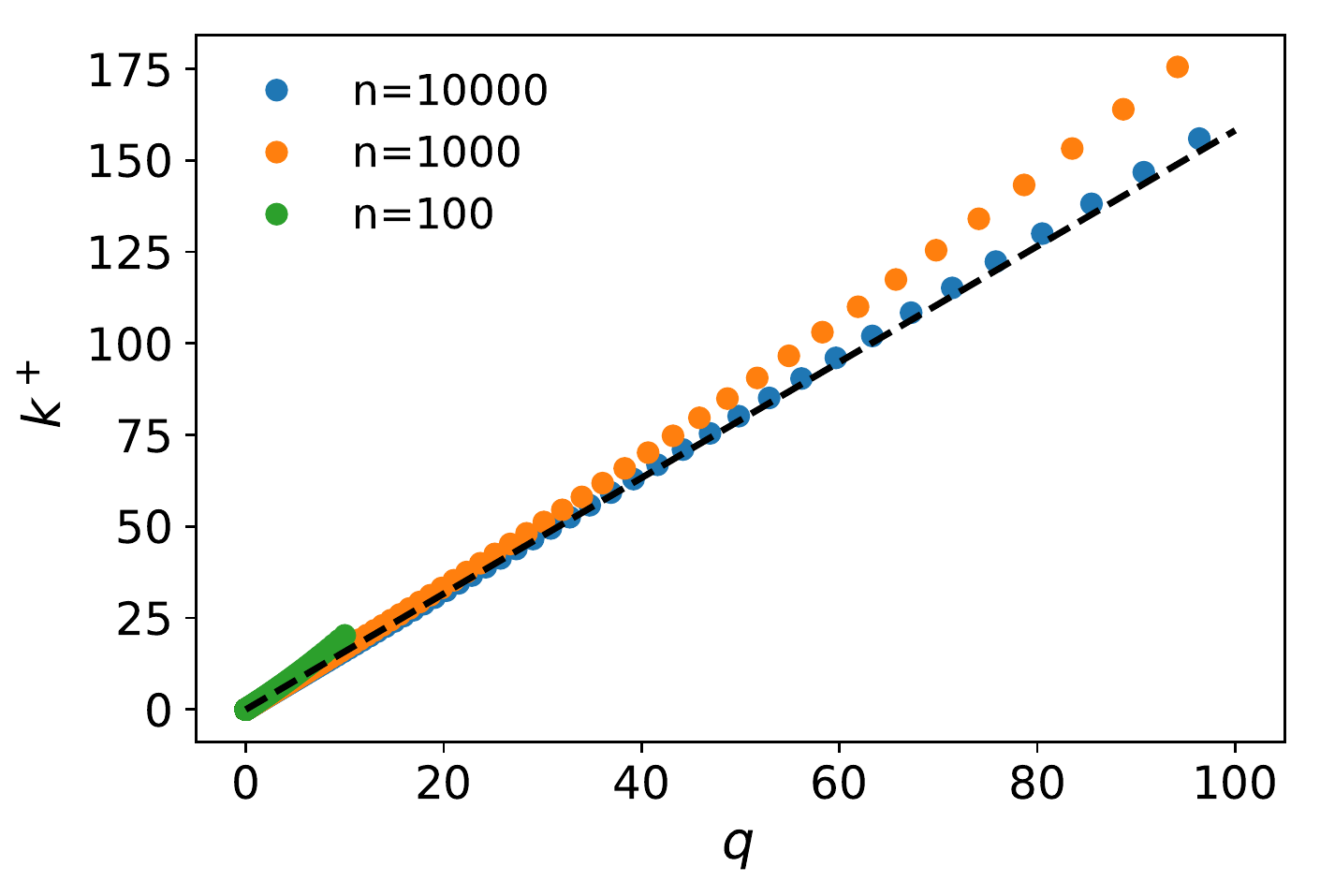} 
\caption{The left plot shows a color map of the analytical value of $k^+(a)$, with $a=1.82$, for different $n$ and $t$. The right plot shows the analytical (dots) and large-system  limit (dotted line) values of $k^+$ obtained for $t$ that span a geometric progression range in $[10,10^4]$ such that $q \in [1,100]$. The large-system limit is obtained from Eq.\eqref{eq:kfa}.}\label{fig:K}
\end{figure}

The above equation suggests to set a threshold $\alpha$ such that $\mathcal{P}(\lambda_0 > 0 )> 1-\alpha$, i.e. $1-\mbox{erf}(a) = \alpha$ (for example, $a \approx 1.82$ for $\alpha=0.01$). One can then define the number of bootstraps required to achieve $\mathcal{P}(\lambda_0 > 0 )> 1-\alpha$ by setting the argument of the erf function to $a$, which gives

\begin{equation}\label{eq:kp}
k^+(a) \simeq \frac{ a^2 \sigma^2(t) + [\mu(t)-1 ] n + \sqrt{a^4 \sigma^4(t) + 2 a^2  \sigma^2(t) [\mu(t)-1] n}}{[\mu(t)-1 ]^2}
\end{equation}
A bi-dimensional mapping of the values of $k^+(a)$ with $a=1.82$ as function of $n$ and $t$, shown in Fig.~\ref{fig:K} left, shows that the number of bootstrap copies $k^+$ required to have a positive defined $\langle \textbf{C} \rangle$ is quite small, at least for not too extreme values of $q=n/t$.\\

To have a rough estimate of the transition point $k^+$ for $\mathcal{P}(\lambda_0>0)\approx 1$ in the limit of large $n$, we can substitute $t = n/q$, and compute the $k^*$ of the inflection point of the error function, obtained for the argument of $\mbox{erf}$ equals to zero 
\begin{equation}\label{eq:flex}
k^* = \frac{2 e n q}{2 (e-1)n +(1 -2 e) q}\,.
\end{equation}
The large-system limit of the first derivative slope at the inflection point if the error function diverges to infinity

\begin{equation}
\lim_{n \to \infty} \frac{d}{d k}\,\left. \frac{k \left[ 2 (e-1) n-2 e q+q\right] -2 e n q}{2  q \sqrt{k \left[ \frac{2 (e-2) n}{q}-e+3\right] } } \right|_{k=k^*} = \infty \,.
\end{equation}

This means that $\mathcal{P}(\lambda_0 > 0)$ has a real transition in the large-system limit. The value of the inflection point of Eq.~\eqref{eq:flex}, in the large-system limit,  converges to
\begin{equation}\label{eq:kfa}
\lim_{n\to \infty} k^*= \lim_{n\to \infty} k^+(a) = \frac{e}{(1-e)}q \approx 1.58\frac{n}{t}
\end{equation} 

The right-hand side of Fig.~\ref{fig:K} shows that this approximation can provide a quite accurate estimation of the magnitude of $k^+$ even for $n$ small when $q$ is not extremely large.

\begin{figure}[thb]
\centering
\includegraphics[width=0.45\columnwidth]{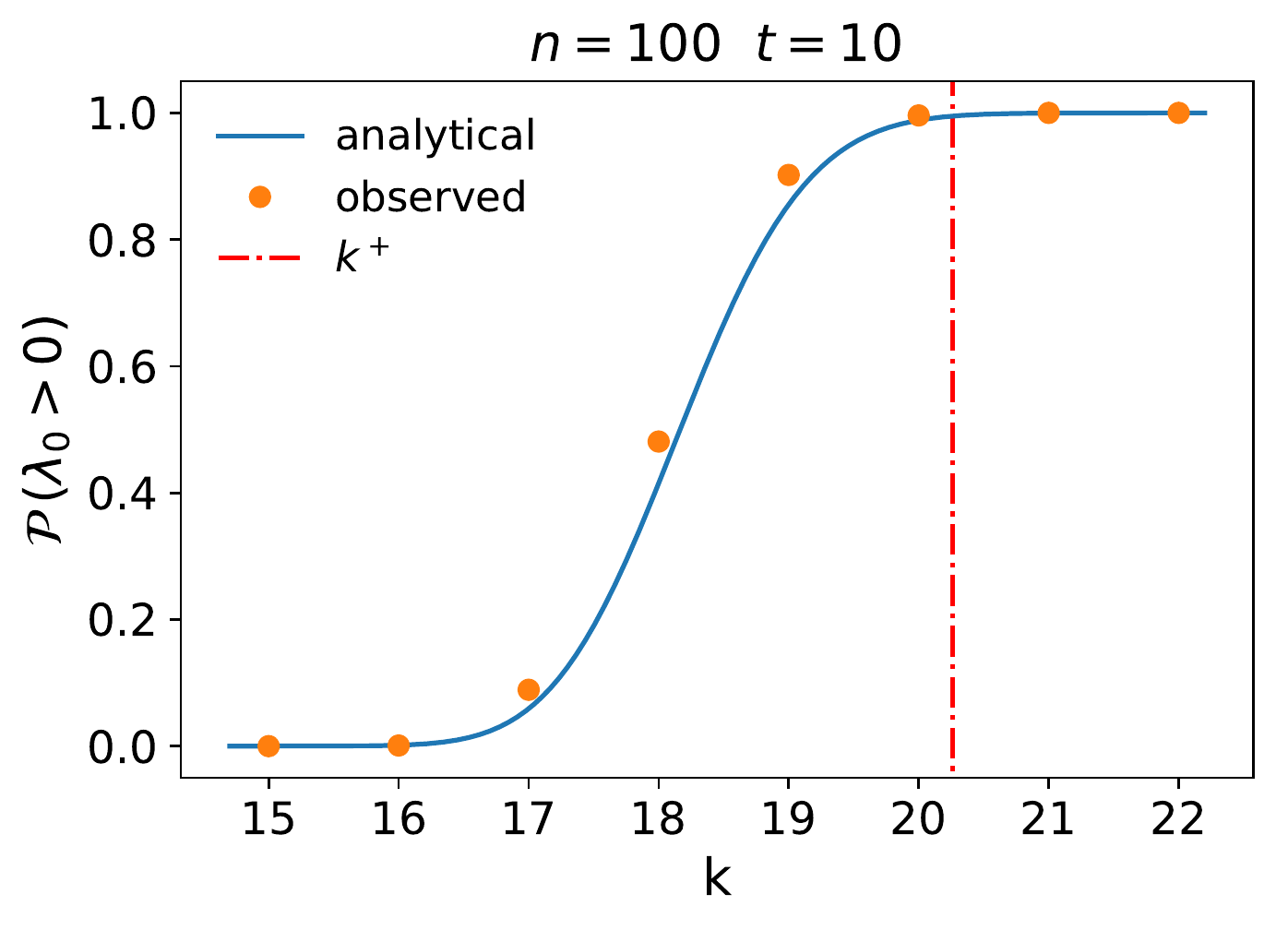} 
\includegraphics[width=0.45\columnwidth]{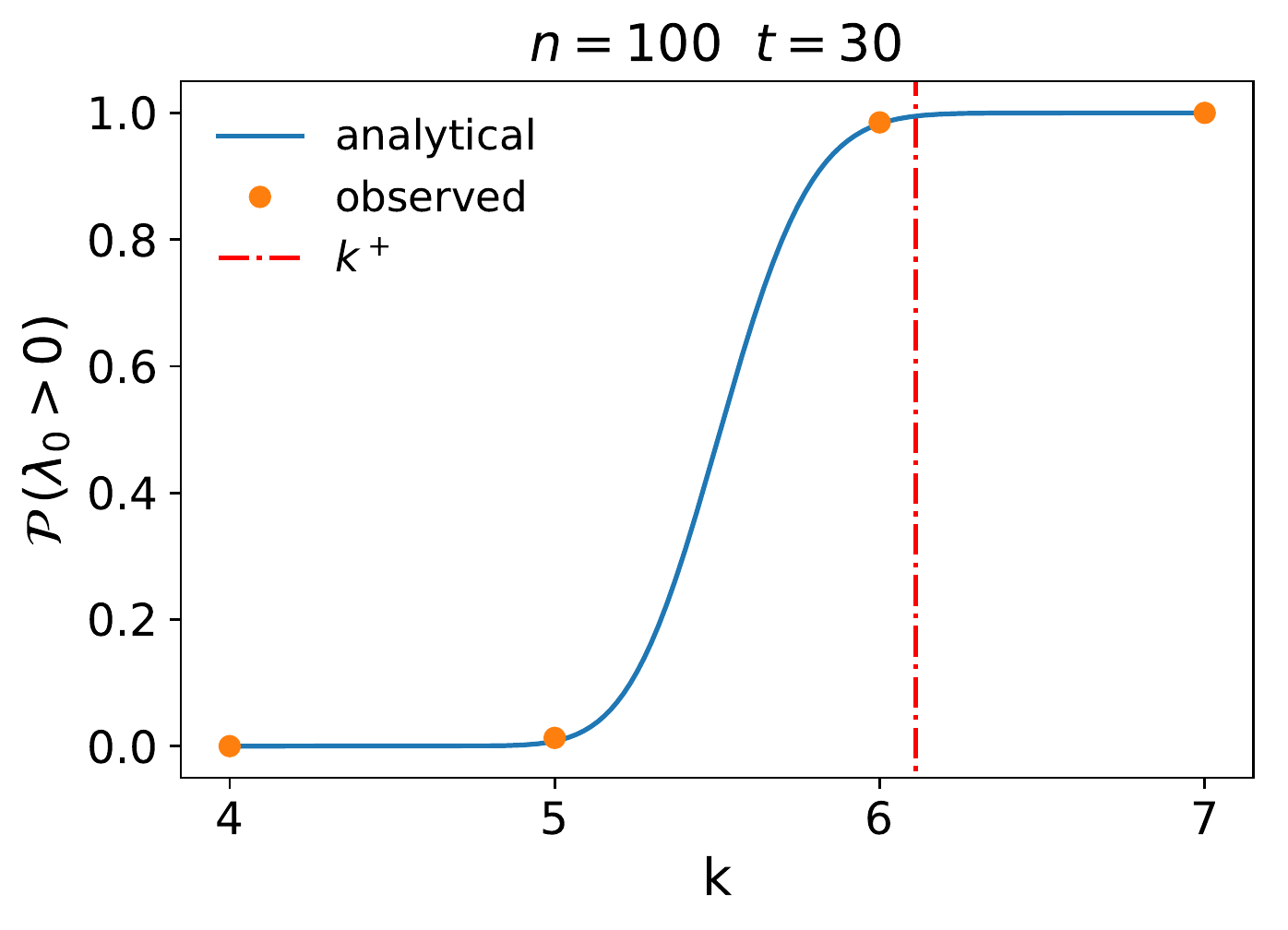} 
\caption{Observed and predicted probability that $\langle \textbf{C} \rangle$ has no zero eigenvalues with $k$ bootstrap copies, for various $n$ and $t$. The figure shows the predicted $k^+(a)$ limit, with $a=1.82$. The simulations are obtained by sampling $\textbf{X}$ from a standardized multivariate normal distribution.}\label{fig:Pl}
\end{figure}

In summary, both the approximate distribution of $\mathcal{P}(\lambda_0>0)$ of Eq.~\eqref{eq:Pl} and the bound limits $k^+$ of Eqs.~\eqref{eq:kp} show a very good agreement with the observations, reported in Fig. \ref{fig:Pl}. 

\section{Discussion}
I have shown that the average correlation matrix of $k$ bootstrap copies converges to a positive-defined matrix for $k$ much smaller than the order of the matrix. Such a matrix can be used in many applications which require to invert $C$, such as risk optimization. An extensive comparative analysis of the performance of these approaches will be addressed in future works.

\section*{Acknowledgements}
I thank prof.~Damien Challet for helpful support and discussions. 
This publication stems from a partnership between CentraleSup\'elec and BNP Paribas.

\section*{References}
\bibliographystyle{unsrt}
\bibliography{report_kbahc}

\begin{thebibliography}{1}

\bibitem{markowitz1959portfolio}
Harry Markowitz.
\newblock {\em Portfolio selection: Efficient diversification of investments},
  volume~16.
\newblock John Wiley New York, 1959.

\bibitem{schafer2005shrinkage}
Juliane Sch{\"a}fer and Korbinian Strimmer.
\newblock A shrinkage approach to large-scale covariance matrix estimation and
  implications for functional genomics.
\newblock {\em Statistical Applications in Genetics and Molecular Biology},
  4(1), 2005.

\bibitem{velasco2015comparative}
Santiago Velasco-Forero, Marcus Chen, Alvina Goh, and Sze~Kim Pang.
\newblock Comparative analysis of covariance matrix estimation for anomaly
  detection in hyperspectral images.
\newblock {\em IEEE Journal of Selected Topics in Signal Processing},
  9(6):1061--1073, 2015.

\bibitem{ledoit2004well}
Olivier Ledoit and Michael Wolf.
\newblock A well-conditioned estimator for large-dimensional covariance
  matrices.
\newblock {\em Journal of Multivariate Analysis}, 88(2):365--411, 2004.

\bibitem{ledoit2017nonlinear}
Olivier Ledoit and Michael Wolf.
\newblock Nonlinear shrinkage of the covariance matrix for portfolio selection:
  Markowitz meets goldilocks.
\newblock {\em The Review of Financial Studies}, 30(12):4349--4388, 2017.

\bibitem{higham2002computing}
Nicholas~J Higham.
\newblock Computing the nearest correlation matrix—a problem from finance.
\newblock {\em IMA Journal of Numerical Analysis}, 22(3):329--343, 2002.

\bibitem{mendelson2016distribution}
Alex~F Mendelson, Maria~A Zuluaga, Brian~F Hutton, and S{\'e}bastien Ourselin.
\newblock What is the distribution of the number of unique original items in a
  bootstrap sample?
\newblock {\em arXiv preprint arXiv:1602.05822}, 2016.

\end{thebibliography}

\end{document}